\documentclass[aps,superscriptaddress,twocolumn,twoside,floatfix,pra,nofootinbib,a4paper]{revtex4-2}
\usepackage{enumerate,appendix}
\usepackage{amsmath, amsthm, amssymb,commath}
\usepackage{color,calc,graphicx}
\usepackage[usenames,dvipsnames,svgnames,table,cmyk,hyperref]{xcolor}
\usepackage[colorlinks]{hyperref}
\hypersetup{
	colorlinks = true,
	urlcolor = {blue},
	citecolor = {magenta},
	linkcolor= {blue}
}

\usepackage{graphicx}
\usepackage{amsmath}
\usepackage{latexsym}
\usepackage{bbm}
\usepackage{soul}

\newcommand{\ket}[1]{|#1\rangle}

\newcommand{\braket}[2]{\langle#1|#2\rangle}

\newcommand{\ketbra}[2]{|#1\rangle\langle#2|}
\usepackage{physics}

\usepackage[charter,cal=cmcal,sfscaled=false]{mathdesign}
\usepackage{booktabs}
\usepackage{multirow}
\usepackage{subfigure}
\usepackage{dcolumn}
\usepackage{mathrsfs}
\usepackage{diagbox}
\usepackage{array}
\usepackage{makecell}
\usepackage{threeparttable}

\begin{document}


\title{Bound entangled states are useful in prepare-and-measure scenarios}

\author{Carles Roch i Carceller}
\affiliation{Physics Department and NanoLund, Lund University, Box 118, 22100 Lund, Sweden.}

\author{Armin Tavakoli}
\email{armin.tavakoli@teorfys.lu.se}
\affiliation{Physics Department and NanoLund, Lund University, Box 118, 22100 Lund, Sweden.}


\begin{abstract}
We show that bipartite bound entangled states make possible violations of correlation inequalities in the prepare-and-measure scenario. These inequalities are satisfied by all classical models as well as  by all quantum models that do not feature entanglement. In contrast to the known Bell inequality violations from bound entangled states, we find that the violations in the prepare-and-measure scenario are sizeable and significantly noise-tolerant. Furthermore, we show that significantly stronger quantum correlations are made possible by considering bound entanglement with a larger dimension. 
\end{abstract}

\maketitle


\textit{Introduction.---} Bound entanglement is a fascinating phenomenon in the theory of quantum entanglement. These are entangled states that, even when available in infinitely many copies, cannot be transformed to a maximally entangled state via local operations and classical communication \cite{Horodecki1998}. In other words, these states are not useful for entanglement distillation. In bipartite systems, it is standard to detect them by verifying that an entangled state has positive partial transpose (PPT). Whether there  exists also bound entanglement that is not PPT is a famous open problem \cite{Horodecki2022}. 

Motivated by understanding the fundamentals of quantum theory, it has for long been asked whether bound entangled states can give rise to various quantum phenomena. For instance, they do support quantum metrology \cite{Toth2018} but not quantum teleportation \cite{Horodecki1999}, albeit some variations of the task are possible for every entangled state \cite{Masanes2006b, Cavalcanti2017}. A more fundamental question is whether bound entanglement can violate principles of classicality in black-box correlation experiments. For example, when entangled states are available in many copies, distillability is necessary and sufficient for violating the CHSH Bell inequality \cite{Masanes2006}.  In 1999 it was conjectured that bound entangled states cannot violate any Bell inequality at all \cite{Peres1999} but this was disproved 15 years later through a counter-example based on two qutrits \cite{Vertesi2014}. The reported violation is strikingly small; already an isotropic noise  rate on fractions of a per mille is sufficient to render the experiment classical. Also subsequent works have  found only small violations of Bell tests with bound entanglement \cite{Vertesi2012, Yu2017, Pal2017, Tendick2020}. This suggests that the original conjecture may have been ``almost true''.

A complementary line of research focuses on the prepare-and-measure scenario. Here, one is interested in the correlations created between Alice and Bob when they share entanglement and communicate messages \cite{Tavakoli2021, Pauwels2022b}. If the message is classical, it is well-known that entanglement enables non-classical correlations \cite{Buhrman2010} but also that this advantage fundamentally is propelled by Bell nonlocality \cite{Brukner2004, Pauwels2022}.  Consequently, if bound entanglement is a resource, it remains limited by the need for significant Bell inequality violations.  However, the situation changes when the messages themselves can be quantum states. This is famously exemplified by the dense coding protocol \cite{Bennett1992}. In dense coding, qubit messages have no advantage over classical bit messages unless also shared entanglement is made available. Many entangled states that violate no Bell inequality can be used for a quantum advantage  in dense coding \cite{Tavakoli2018, Moreno2021}, but it turns out that bound entangled states are not among these \cite{PrivateCommunication}.

Here, we show that bound entangled states of two particles make possible  quantum correlations in the prepare-and-measure scenario that significantly violate classical limitations. Specifically, we show large violations of correlation inequalities that are satisfied by all classical models and  all quantum models that do not exploit entanglement. This shows not only that bound entanglement is responsible for breaking the  classical limitations but also that its advantages can be sizable and therefore also of practical relevance. In our analysis, we consider the black-box input-output scenario based on $d$-dimensional messages recently introduced in \cite{Bakhshinezhad2024}. Using semidefinite relaxation techniques \cite{Tavakoli2024}, we prove a relevant correlation inequality in this scenario. Then, through an explicit construction, we show that it can be violated with  bound entangled states of two qutrits. As a benchmark of the violation, the  bound entanglement remains non-classical when exposed to up to $18.8\%$ of  isotropic noise, which far exceeds the known counterparts in bipartite Bell scenarios. We subsequently consider correlation inequalities tailored for systems with a higher Hilbert space dimension and prove that even more noise-tolerant quantum correlations are now made possible by using bound entangled states. Our results establish bound entanglement as useful for the prepare-and-measure scenario in both principle and practice. \\

\begin{figure}[t!]
	\centering
	\includegraphics[width=\columnwidth]{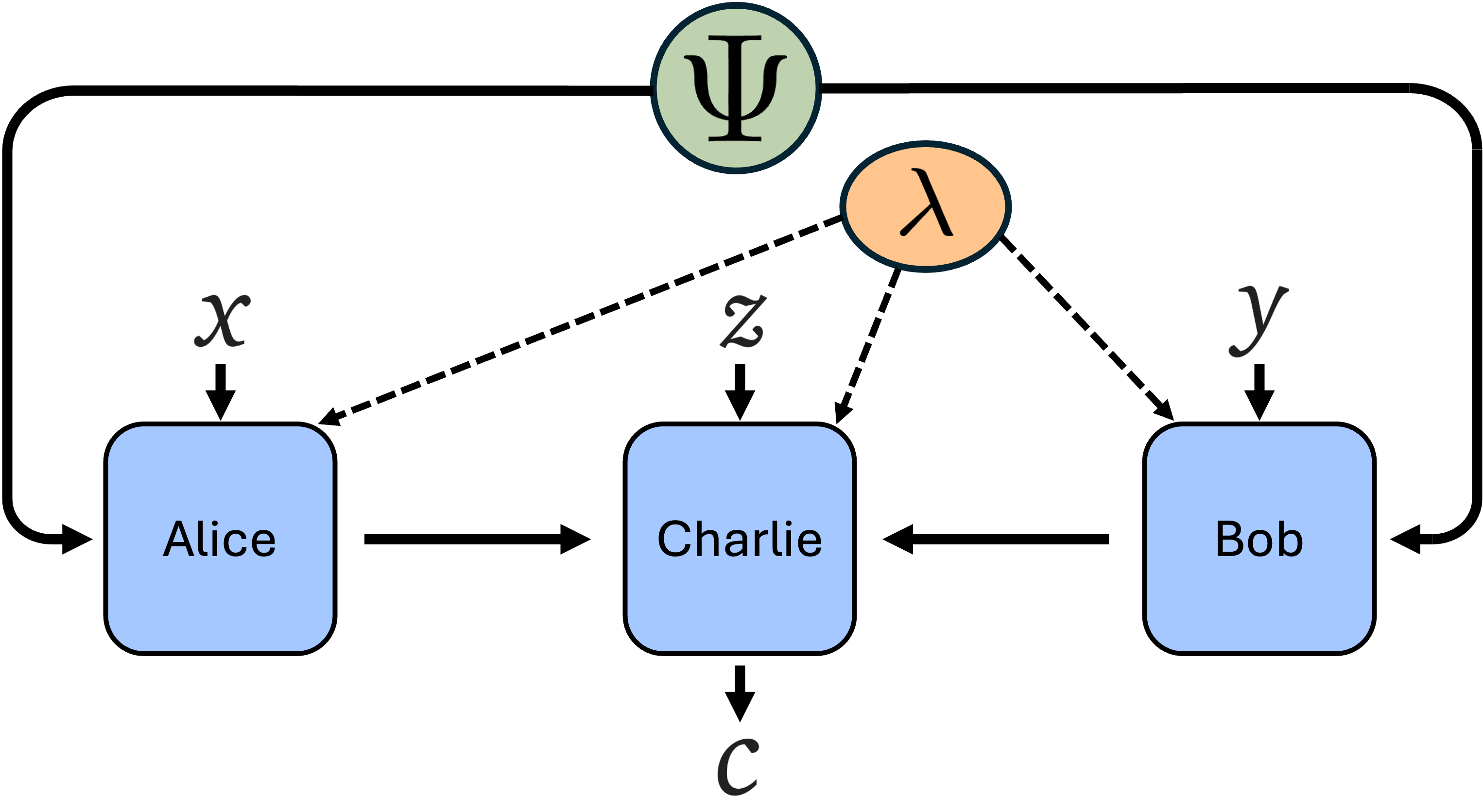}
	\caption{Symmetric prepare-and-measure scenario. All parties share classical randomness ($\lambda$). When entanglement is allowed, any quantum state $\Psi$ may be shared between Alice and Bob. They respectively encode their inputs $x$ and $y$ into $d$-dimensional messages that are sent to Charlie who decodes them with respect to his input $z$ and outputs $c$.}\label{FigScenario}
\end{figure}

\textit{Scenario.---} We consider the symmetric prepare-and-measure scenario illustrated in Figure~\ref{FigScenario}. Alice, Bob and Charlie select private inputs $x$, $y$ and $z$ respectively. Only Charlie produces an output, and it is labeled $c$. Alice and Bob can each send a message to Charlie and these messages support at most $d$ distinct symbols. Thus, when the message is a quantum system, it corresponds to a density matrix with Hilbert space dimension $d$. Moreover, all three parties share unlimited classical randomness, corresponding to a hidden variable, $\lambda$, subject to some arbitrary probability distribution $\{q_\lambda\}$. This can be used to stochastically coordinate the communication strategies between the three parties. Also, Alice and Bob can share some arbitrary entangled state, $\Psi$, to assist their respective selection of  messages. The correlations obtained between the three parties, $p(c|x,y,z)$, are in this entanglement-based model given by 
\begin{equation}
	p_E(c|x,y,z)=\sum_\lambda q_\lambda \tr\left[\left(\Lambda_{x}^{\lambda}\otimes\Omega_{y}^{\lambda}\right)[\Psi]M_{c|z}^{\lambda}\right],
\end{equation}
where $\{M_{c|z}^\lambda\}$ are the measurements of Charlie, and $\{\Lambda_{x}^\lambda\}$ and $\{\Omega_{y}^\lambda\}$ are the message encoding maps of Alice and Bob respectively, all when conditioned on the hidden variable. Each of the encoding maps corresponds to a completely positive trace-preserving map. These maps transform the arbitrary input space  associated with the respective shares of $\Psi$ into a $d$-dimensional system. 

Consider now a classical model of the same scenario. All three parties can still pre-share the hidden variable but the entangled state, $\Psi$, is no longer available. Also, the communication itself is classical, i.e.~Alice and Bob  map their respective inputs to classical messages $m_A\in[d]$ and $m_B\in[d]$ which are sent to Charlie. We define $[d]\equiv\{0,\ldots,d-1\}$. Charlie then computes his output based on the received messages, his own input $z$ and the hidden variable $\lambda$. Thus, the classical correlations take the form 
\begin{align}\label{classical}
	p_C(c|x,y,z)=\!\!\!\!\sum_{\lambda,m_A,m_B}\!\!\!\! q_\lambda p(m_A|x,\lambda)p(m_B|y,\lambda)p(c|m_A,m_B,z,\lambda).
\end{align} 
The space of classical correlations in prepare-and-measure scenarios takes the geometric form of a polytope \cite{Gallego2010} and its facets correspond to linear correlation inequalities satisfied by all classical models. In general, these inequalities take the form
\begin{equation}\label{ineq}
\sum_{c,x,y,z} s_{cxyz}p(c|x,y,z) \leq \beta,
\end{equation}
for some real coefficients $s_{cxyz}$ and some tight bound $\beta$. Thus, violating such an inequality implies non-classical correlations in the prepare-and-measure scenario. However,  a violation is insufficient to guarantee that entanglement is the essential resource behind the failure of classicality. In order to guarantee the role of entanglement, one must also show that the inequality \eqref{ineq} is satisfied also by all quantum models that do not use entanglement, i.e.~when the hidden variable is used for coordinating the communication of quantum messages to Charlie. The correlations in such an ``unentangled'' quantum model take the form 
\begin{equation}\label{quantum}
	p_Q(c|x,y,z)=\sum_\lambda q_\lambda \tr\left[\left(\rho_{x}^\lambda\otimes \sigma_{y}^\lambda \right)M_{c|z}^\lambda\right],
\end{equation} 
where $\{\rho_{x}^\lambda\}$ and $\{\sigma_{y}^\lambda\}$ are the $d$-dimensional quantum messages of Alice and Bob when conditioned on the hidden variable. The set of correlations that admit this model is convex and constitutes, in general, a strict superset of the classical polytope. The latter follows from the fact that quantum messages can lead to non-classical correlations in the prepare-and-measure scenario \cite{Wiesner1983}. In what follows, we proceed to  construct specific correlation inequalities of the form \eqref{ineq} which has the key feature that they are satisfied  by both classical models \eqref{classical} and  quantum models without entanglement \eqref{quantum}. \\

\textit{Correlation inequality.---} We consider the input-output scenario introduced in \cite{Bakhshinezhad2024}. It is parameterised by $d$, with $x\equiv (x_0,x_1)\in[d]^2$, $y\equiv (y_0,y_1)\in[d]^2$,  $z\in[d+1]$ and $c\in[d]$. Each $z$ corresponds to a game that has a unique right winning answer, $w_z(x,y)$, and Charlie's aim is to output $c=w_z(x,y)$. We define the winning answers as 
\begin{align}\nonumber
& z\neq d: && w_z=x_1+y_1-2z(x_0-y_0) \mod{d}\\
& z= d: && w_d=x_0-y_0 \mod{d}
\end{align}
and define the average winning probability as
\begin{equation}
\mathcal{R}_d=\frac{1}{d^4(d+1)}\sum_{x,y,z}p(c=w_z(x,y)|x,y,z).
\end{equation}
While this definition applies to arbitrary $d$, our focus will be on prime $d$ due to the favourable property of there always existing multiplicative inverses for the winning conditions. This is expected to be useful for finding entanglement-based advantages. It was conjectured in \cite{Bakhshinezhad2024} that for any quantum model without shared entanglement and every odd prime $d$ it holds that 
\begin{equation}\label{ourineq}
\mathcal{R}_d\leq \frac{2}{d+1}.
\end{equation}
It is easy to see that this conjectured bound can be saturated also in a classical model. Let Alice and Bob send $m_A=x_0$ and $m_B=y_0$ to Charlie. When $z=d$, Charlie will correctly answer $c=w_d$ and when $z\neq d$ he can make a random guess which has winning probability $p(c=w_z)=1/d$. This leads to $\mathcal{R}_d=2/(d+1)$. Thus, if the conjecture is true, Eq.~\eqref{ourineq} is an instance of a correlation inequality \eqref{ineq} that is tight for both for classical models and quantum models without entanglement.

We have proven the conjecture for $d=3,5,7$. The proof is based on identifying specific super-sets of the set of quantum correlations \eqref{quantum} achievable without entanglement, and then showing that the conjecture \eqref{ourineq} is true also for  these sets. The super-sets are constructed so that they can be characterised as semidefinite programs. Such programs can be solved exactly, and via duality  theory they can also provide analytical certificates of the result \cite{Boyd2004, Tavakoli2024}.
	
To this end, we define a tracial moment matrix with elements $\Gamma_{u,v}=\tr(u^\dagger v)$ where $u,v\in L_k$ are monomials up to a selected degree $k$ over the operator list $\mathcal{S}=\{\Pi_A\otimes \Pi_B,\{\rho_x\otimes\openone\},\{\openone\otimes \sigma_y\},\{M_{c|z}\}\}$. Here, we use the idea of Refs.~\cite{Gribling2018, Pauwels2022c, Alessandro2024} to assign to $\Pi_A$ and $\Pi_B$ the properties of a $d$-dimensional projector, i.e.~we impose that $\Pi^2=\Pi$ and $\tr(\Pi)=d$. Moreover, the states are fully contained in the support of these projectors, i.e.~$\Pi_A\rho_x=\rho_x$ and $\Pi_B\sigma_y=\sigma_y$. We can also w.l.g.~assume that the states are pure because the classical randomness in a mixed state can be absorbed into the global hidden variable $\lambda$. That $\rho$ and $\sigma$ live on separate spaces is captured through their commutation.  Further, we assume the measurements to be projective i.e.~$M_{c|z}M_{c'|z}=\delta_{cc'}M_{c|z}$. These conditions together with the cyclicity of the trace imply many constrains on the moment matrix $\Gamma$. By construction, we have $\Gamma\succeq 0$ and the functional $\mathcal{R}_d$ can be expressed as a linear combination over the elements of $\Gamma$, provided the degree $k$ of the monomials is sufficiently high. Thus, we can compute an upper bound on $\mathcal{R}_d$ as a semidefinite program. We have specifically selected monomials of length $k=1$ supplemented with the length $k=2$ monomials corresponding to the products $\{(\rho_x\otimes\openone)M_{c|z}\}$. 

However, the  complexity of this semidefinite program scales rapidly with growing dimension, making it too costly for us to solve beyond the simplest relevant case of $d=3$. This can be remedied by using symmetrisation techniques  to  simplify the structure of $\Gamma$ (see e.g.~\cite{Gatermann2004, Tavakoli2019, Ioannou2022}). For this, we observe that there exists local permutations of the variables of Alice, Bob and Charlie that leave $\mathcal{R}_d$ invariant. By analysing these permutations, one can reduce both the number of variables in $\Gamma$ and its size. This is detailed in Supplementary Material. Showcasing the relevance of this method, we have evaluated it for $d=3,5,7$ and obtained the inequality \eqref{ourineq}. To obtain a certificate of the result, we have also derived the dual program (see Supplementary Material) and confirmed that it gives the accurate result.

\textit{Violation with bound entanglement.---} We now show that the inequality \eqref{ourineq} can be violated in the simplest case of $d=3$ by using bound entanglement. For this purpose, we will consider a specific bound entangled state of  two qutrits. Its density matrix takes the form 
\begin{equation}\label{boundent}
\rho_{\text{BE}}=\begin{pmatrix}
a_1 & 0 & 0 &0 & a_2 & 0 & 0 & 0 &a_2^*\\
0 & a_3 & 0 & 0 & 0 &a_4 & a_4^* &0 & 0\\
0&0&a_1&a_5&0&0&0&a_5^*&0\\
0&0&a_5^*&a_1&0&0&0&a_5&0\\
a_2^*&0&0&0&a_1&0&0&0&a_2\\
0&a_4^*&0&0&0&a_3&a_4&0&0\\
0&a_4&0&0&0&a_4^*&a_3&0&0\\
0&0&a_5&a_5^*&0&0&0&a_1&0\\
a_2&0&0&0&a_2^*&0&0&0&a_1
\end{pmatrix},
\end{equation}
where $(a_1,a_3)$ are real numbers and $(a_2, a_4,a_5)$ are complex numbers. We select these numbers as
\begin{align}\nonumber
& a_1=\frac{4-\sqrt{3}}{27},  \qquad a_2=\frac{1-\sqrt{3}}{27}-\frac{i}{9\sqrt{3}}, \qquad a_3=\frac{1+2\sqrt{3}}{27}, \\
 & a_4=\frac{5-2\sqrt{3}}{54}+i\frac{5\sqrt{3}-12}{54},	\qquad a_5=\frac{\sqrt{3}-4}{54}+\frac{i}{18}.
\end{align}
Given these choices, one can easily verify that $\rho_{\text{BE}}$ has unit trace and that it is positive semidefinite. Its rank is seven and its non-zero eigenvalues with their corresponding multiplicities are  $\{\frac{\sqrt{3}-1}{3},(\frac{2}{9})^{\otimes 3},(\frac{2-\sqrt{3}}{9})^{\otimes 3}\}$. We also see that $\rho_\text{BE}$ has PPT. The spectrum of  $\rho_\text{BE}^{\text{T}_A}$ is $\{(\frac{4-\sqrt{3}}{9})^{\otimes 3}, (\frac{\sqrt{3}-1}{9})^{\otimes 3}, 0^{\otimes 3}\}$ which is non-negative. The fact that $\rho_\text{BE}$ is entangled can be certified from the computable cross norm or realignment criterion. To see this, let $\{\gamma_i\}_{i=1}^{d^2}$ be an orthonormal basis of $d$-dimensional Hermitian matrices, i.e.~$\tr(\gamma_j\gamma_k)=\delta_{j,k}$. The criterion states that for all separable states $\|C\|_\text{tr}\equiv \tr\sqrt{C^\dagger C}\leq 1$ where $C_{j,k}=\tr(\gamma_j\otimes\gamma_k \rho)$ \cite{Chen2003, Rudolph2005}. The violation obtained from choosing $\rho=\rho_\text{BE}$ is $\|C\|_\text{tr}=\frac{1}{3}\left(2\sqrt{3}+\sqrt{6}-\sqrt{2}-1\right)\approx 1.166$. 

Next, we use $\rho_{\text{BE}}$ to violate the inequality \eqref{ourineq}. We select Alice's and Bob's encoding maps as unitary transformations. Specifically, we take $\Lambda_x(\tau)= U_x\tau U_x^\dagger$ with $U_x=X^{x_0}Z^{x_1}$ where $X=\sum_{k=0}^{d-1}\ketbra{k+1}{k}$ and $Z=\sum_{k=0}^{d-1}\omega^{k}\ketbra{k}{k}$ are the shift and clock operators, with $\omega=e^{\frac{2\pi i}{d}}$. We choose Bob's encoding maps as the same unitaries, namely $\Omega_y=\Lambda_y$. Note that we do not exploit the hidden variable and hence we drop the label for $\lambda$. Finally, we must select the measurements of Charlie. To this end, we define the bases
\begin{equation}
\ket*{e_{l|j}}=\frac{1}{\sqrt{d}}\sum_{k=0}^{d-1} \omega^{kl+jk^2}\ket{k},
\end{equation}
where $l\in[d]$ is the outcome label and $j\in[d]$ is the basis label. Select also $\ket{e_{l|d}}=\ket{l}$  (i.e.~$j=d$) as the computational basis. For any prime $d$, the $d+1$ bases associated with $j$ are mutually unbiased, i.e.~it holds that $|\braket*{e_{l|j}}{e_{l'|j'}}|^2=\frac{1}{d}$ when $j\neq j'$ \cite{Wootters1989}. For simplicity, we write $E_{l|j}=\ketbra*{e_{l|j}}{e_{l|j}}$. Now, for the case of $d=3$, we use these bases to construct the measurements of Charlie as 
\begin{equation}
M_{c|z}=\sum_{b=0}^{d-1} E_{b|z}\otimes \tilde{E}_{b-c|z}.
\end{equation}
Here, we have defined $\tilde{E}$ as a relabeling of the bases $E$ and their eigenvectors. Specifically, we choose the tuple $\left(\tilde{E}_{0|z}, \tilde{E}_{1|z}, \tilde{E}_{2|z}\right)$ as follows: $\left(E_{1|0},E_{0|0},E_{2|0}\right)$ for $z=0$,  $\left(E_{0|2},E_{2|2},E_{1|2}\right)$ for $z=1$,  $\left(E_{1|1},E_{0|1},E_{2|1}\right)$ for $z=2$ and  $\left(E_{1|3},E_{2|3},E_{0|3}\right)$ for $z=3$.  Evaluating the figure of merit, we obtain the value
\begin{equation}\label{violation}
\mathcal{R}_3=\frac{1}{4}+\frac{1}{2\sqrt{3}}\approx 0.5387 >\frac{1}{2}.
\end{equation} 
This is a violation of the inequality \eqref{ourineq}. For completeness, we note that, as expected, the violation is much smaller than what is possible with a maximally entangled state. In that case, one can achieve $\mathcal{R}_d=1$ for prime $d$. \cite{Bakhshinezhad2024}.

Maximally entangled states provide some insight as to why bound entanglement can achieve the reported violation \eqref{violation}. The connection stems from the observation that the  $\rho_\text{BE}$ is a Bell-diagonal state, i.e.~that it can be written on the form $\rho_\text{BE}=\sum_{k=1}^9 q_k \ketbra*{\psi_k}{\psi_k}$ where the mixing coefficients $q_k$ correspond to the eigenvalues of $\rho_\text{BE}$ and $\{|\psi_k\rangle\}$ is the Bell basis (i.e.~the standard basis of maximally entangled two-qutrit states). To better understand the violation, we compute the value of $\mathcal{R}_3$ for each of the states $|\psi_k\rangle$. Among the nine states, two give $\mathcal{R}_3=0$, three give $\mathcal{R}_3=\frac{1}{4}$, another three give $\mathcal{R}_3=\frac{1}{2}$ and the final one gives $\mathcal{R}_3=\frac{3}{4}$. The first correspond to the null-eigenvalues of $\rho_\text{BE}$, the second to the smallest non-zero eigenvalue, the third to the second largest eigenvalue and the final to the largest eigenvalue. Thus, we see that the violation is made possible because of the skewed distribution of eigenvalues $\{q_k\}$ which is nevertheless compatible with PPT. It lets us assign only small (or even zero) eigenvalues to choices of $\ket*{\psi_k}$ that achieve $\mathcal{R}_3<\frac{1}{2}$, while it assigns a relatively large eigenvalue to the specific $\ket*{\psi_k}$ that achieves a sizable violation. This leads to an average over the Bell basis, $\mathcal{R}_3=\sum_k q_k \mathcal{R}_3(\ket*{\psi_k})$, which is sufficient to achieve \eqref{violation}.

Importantly, we also want to quantify how far the quantum correlations are from satisfying the inequality. This can be done in many different ways, but a common benchmark is to consider the robustness when the shared state is exposed to isotropic noise. This amounts to considering the density matrix $\rho_\nu=\frac{\nu}{d} \openone+(1-\nu)\rho_\text{BE}$, where $\nu\in[0,1]$ is the noise rate. Evaluating $\mathcal{R}_3$, we find $\mathcal{R}_3=\frac{1}{12}\left(3+2\sqrt{3}-(2\sqrt{3}-1)\nu\right)$. The critical $\nu$ for achieving a violation becomes 
\begin{equation}
\nu=\frac{9-4\sqrt{3}}{11}\approx 0.1883.
\end{equation}
This indicates a sizable robustness of the non-classicality generated by the bound entangled state. 

We remark that we have also performed numerical search over bound entangled states, the encodings of Alice and Bob and the measurements of Charlie, and we have found no larger violation of the inequality. \\

\begin{table}[t!] 
	\begin{tabular}{c|ccc}
		\cellcolor[HTML]{FFFFFF}- & \begin{tabular}[c]{@{}c@{}}$\mathcal{R}_d$\\ w/o Ent\end{tabular} & \begin{tabular}[c]{@{}c@{}}$\mathcal{R}_d$\\ with bound Ent\end{tabular} & \begin{tabular}[c]{@{}c@{}}Isotropic \\ noise-tolerance\end{tabular} \\ \hline
		$d=3$                     & $1/2$                                                     & $0.5387$                                                                 & $0.1883$                                                             \\
		$d=5$                     & $1/3$                                                     & $0.3862$                                                                 & $0.2839$                                                             \\
		$d=7$                     & $1/4$                                                     & $0.2931$                                                                      & $0.2869$                                                               
	\end{tabular}
	\caption{Violations of inequality \eqref{ourineq} using bound entangled states of dimension $d=3,5,7$. For the optimal bound entangled state, we consider also its mixture with isotropic noise and determine the maximal noise rate for which a violation is possible.}\label{Tab1}
\end{table}

\textit{Higher dimensional violations.---} Given that large violations of classicality are possible in the prepare-and-measure scenario using three-dimensional bound entanglement, it appears natural to ask whether considering higher dimensions will lead to even stronger quantum correlations. This is a priori not obvious since several inequalities for the prepare-and-measure scenario are known to have diminishing quantum violations with growing dimension 
\cite{Brunner2013, Tavakoli2015}. Interestingly, we find that for $d=5$ and $d=7$ even larger advantages are possible from bound entangled states than we found in dimension $d=3$. To show this, we have employed an alternating convex search method \cite{Tavakoli2024} in which we set the encoding maps of Alice and Bob to the Weyl-Heisenberg unitaries, namely $\{X^{j}Z^{k}\}_{j,k=0}^{d-1}$, and consider optimisation over the state and the measurements of Charlie. Each of these two optimisations can be written as a semidefinite program over matrices of dimension $d^2$, and we repeatedly iterate between the two in order to maximise the value of $\mathcal{R}_d$. We discuss this procedure in detail in Supplementary Material. The results are given in Table~\ref{Tab1}. We see that the noise-robustness improves significantly compared with the case of $d=3$. \\

\textit{Conclusions.---} We have shown that bound entanglement is a resource for non-classical correlations in the prepare-and-measure scenario, and that it can exhibit significant levels of noise tolerance. The noise tolerance exceeds that known for Bell nonlocality from bipartite bound entanglement by orders of magnitude. This noise-robustness makes our schemes viable from an implementation perspective. Moreover, while most schemes in the entanglement-assisted prepare-and-measure scenario employ entangled measurements, ours uses only product measurements \cite{Piveteau2022, Piveteau2024} to reveal the role of bound entanglement. This is many times a considerable simplification in the complexity of the measurement.  Our results for $d=3,5,7$ showcase a significantly increasing noise-tolerance for quantum correlations from bound entanglement. It motivates a question for future work: can bound entanglement make possible a diverging correlation advantage? Specifically, in the limit of large dimension, can bound entanglement break the limitations of both classical models and  quantum models without entanglement, when subject to noise rates that tend to one for large dimensions?  \\

\begin{acknowledgments}
We thank Pharnam Bakhshinezhad, Mohammad Mehboudi, Lucas Tendick and Tam\'as V\'ertesi for discussions. We thank Paul Skrzypczyk and Chung-Yun Hsieh for sharing their notes on dense coding. This work is supported by the Wenner-Gren Foundations, by the Knut and Alice Wallenberg Foundation through the Wallenberg Center for Quantum Technology (WACQT) and the Swedish Research Council under Contract No.~2023-03498. \\
\end{acknowledgments} 
	
\textit{Code availability}: The code used to prove the conjecture in \eqref{ourineq} for $d=3,5,7$ is available on GitHub: \url{https://github.com/chalswater/bound_entanglement_conjecture}.
	
\bibliography{references_ppt}

\appendix

\begin{widetext}

\newpage

\section{Alternating convex search method}
We detail our search method for optimising $\mathcal{R}_d$ over bound entangled states. Ours is an interior-point search, meaning that we search over explicit quantum strategies. Thus, the solution returns a specific bound entangled state,  measurements and channels. Therefore, any violation returned by this method can be analytically verified.

In a quantum model with entanglement, the value of $\mathcal{R}_d$ is given by
\begin{equation}
\mathcal{R}_d=\frac{1}{d^4(d+1)}\sum_{x,y,z} \tr\left[\left(\Lambda_{x}\otimes\Omega_{y}\right)[\Psi]M_{w_z|z}\right]
\end{equation}
where $\{\Lambda_x\}$ are the channels of Alice, $\{\Omega_y\}$ are the channels of Bob, $\{M_{c|z}\}$ are the measurements of Charlie and $\Psi$ is the (bound) entangled state shared initially between Alice and Bob.  Note that we have ignored shared randomness because the optimal value of any linear expression (like $\mathcal{R}_d$) is achieved deterministically. Shared randomness can therefore safely be ignored.

We  select Alice's and Bob's  channels  as the unitary generators of the Weyl-Heisenberg group. Specifically, we take Alice's channels to be $\Lambda_x(\tau)= U_x\tau U_x^\dagger$ with $U_x=X^{x_0}Z^{x_1}$ where $X=\sum_{k=0}^{d-1}\ketbra{k+1}{k}$ and $Z=\sum_{k=0}^{d-1}\omega^{k}\ketbra{k}{k}$ are the shift and clock operators, with $\omega=e^{\frac{2\pi i}{d}}$. We choose Bob's channels to be identical,  namely $\Omega_y=\Lambda_y$. Thus, we are left with optimising over the measurements of Charlie and the bound entangled state.

Consider now that we are given some arbitrary measurement for Charlie, $\{M_{c|z}\}$. The  optimal value of $\mathcal{R}_d$ over the bound entangled state $\Psi$ is written as
\begin{align}\label{sdp1}
\underset{\Psi}{\text{maximize}} & \quad \mathcal{R}_d\\
\text{subject to} & \quad \tr(\Psi)=1, \quad \Psi^{T_A}\succeq 0, \quad \Psi\succeq 0. \nonumber
\end{align}
Here, bound entanglement is imposed through the positive partial-transpose condition. This defines a semidefinite program because $\mathcal{R}_d$ is linear and the domain of optimisation is positive semidefinite matrices subject to linear equalities. Hence, it can be evaluated using standard means \cite{Boyd2004}.

Consider now that we are given a bound entangled state $\Psi$. We can then compute the optimal value of $\mathcal{R}_d$ as a semidefinite program over the measurements $\{M_{c|z}\}$. This program is written as
\begin{align}\label{sdp2}
\underset{\{M_{c|z}\}}{\text{maximize}} & \quad \mathcal{R}_d\\\nonumber
\text{subject to} & \quad \sum_{c}M_{c|z}=\openone \ \forall z, \quad M_{c|z}\succeq 0 \ \forall c,z. \nonumber
\end{align}

Thus, in order to search for the optimal value of $\mathcal{R}_d$ over generic bound entangled states and generic measurements, we can iteratively implement the semidefinite programs \eqref{sdp1} and \eqref{sdp2}. That is, we start by drawing a random measurement, then we run the program \eqref{sdp1}. We then use the bound entangled state returned by this program as the input to the program \eqref{sdp2}. The measurement returned by this program then becomes the input to the program \eqref{sdp1}, and so on. This loop is repeated until we see that $\mathcal{R}_d$ converges. We typically repeat the whole procedure several times since it can depend on the initial random choice of the measurement.

\section{SDP relaxation, parameter reduction by symmetrization and dualization}

In this part of the supplemental material we present the semidefinite programming (SDP) relaxation used in the main text to prove the main conjecture.

\subsection{SDP relaxation}

The central scenario presented in the main text is a communication game played in the symmetric prepare-and-measure scenario. Alice and Bob each receive a share of a quantum state $\Psi$ with local dimension $d$ and encode a message of size $d$ using the symbols $x\equiv(x_0,x_1)\in[d^2]$ and $y\equiv(y_0,y_1)\in[d^2]$ respectively. These messages are then sent to Charlie, who performs a decoding scheme using $z\in[d+1]$ distinct measurements and observes $c\in[d]$ as measurement outcomes. This communication game is characterised by the winning condition $w_z=c$, for
\begin{align}\label{eq:winning_condition}
z\neq d: \ w_z=x_1+y_1-2z(x_0-y_0) \mod{d} \ , \quad \qquad z= d: \ w_d=x_0-y_0 \mod{d} \ .
\end{align}
We write the average winning probability as
\begin{equation}
\mathcal{R}_d=\frac{1}{d^4(d+1)}\sum_{x,y,z}p(c=w_z(x,y)|x,y,z) \ .
\end{equation}
Our goal is to find the maximum $\mathcal{R}_d$ achievable using quantum strategies that do not use entanglement. This implies that $\Psi$ is unentangled, allowing us to directly define $\rho_{x_0 x_1}$ and $\sigma_{y_0 y_1}$ as the quantum states in which Alice and Bob encode their messages, respectively. The observable probabilities can then be written using the Born rule as
\begin{align}
p(c|x y z) = \Tr\left[M_{c|z}\left(\rho_{x_0 x_1}\otimes \sigma_{y_0 y_1}\right)\right] \
\end{align}
modulo the use of shared randomness. However, since we aim at maximising the linear functional $\mathcal{R}_d$, the optimum will be achievable deterministically (i.e~without shared randomness) and it can therefore be safely ignored.

To find a bound on $\mathcal{R}_d$ we will employ an SDP relaxation which consists in building a tracial moment matrix. Specifically, we define the list of relevant operators $\mathcal{S}=\{\Pi_A\otimes\Pi_B,\{\rho_{x}\otimes\openone\}_{x},\{\openone\otimes\sigma_{y}\}_{y},\{M_{c|z}\}_{c,z}\}$. From $\mathcal{S}$ we build the list of monomials $L=\{\Pi_A\otimes\Pi_B,\{\rho_{x}\otimes\openone\}_{x},\{\openone\otimes\sigma_{y}\}_{y},\{M_{c|z}\}_{c,z},\{(\rho_{x}\otimes\openone)M_{c|z}\}_{x,c,z}\}$. Here $\Pi_{A/B}$ are projectors onto $d$-dimensional spaces, thereby serving as the identity operator. We use these monomials to build the moment matrix $\Gamma$ with elements $\Gamma_{u,v}=\tr\left(uv^\dagger\right)$, for $u,v \in L$. The moment matrix becomes
\begin{align}
\Gamma = \begin{pmatrix}
\tr(\Pi_A\otimes\Pi_B) & \left[\tr(\rho_{x}\otimes\openone)\right]_x & \left[\tr(\openone\otimes\sigma_{y})\right]_y & \left[\tr(M_{c|z})\right]_{c,z} & \left[\tr((\rho_{x}\otimes\openone)M_{c|z})\right]_{x,c,z} \\
 & \left[\tr(\rho_{x}\rho_{x'}\otimes\openone)\right]_{x}^{x'} & \left[\tr(\rho_{x'}\otimes\sigma_{y})\right]_{y}^{x'} & \left[\tr((\rho_{x'}\otimes\openone)M_{c|z})\right]_{c,z}^{x'} & \left[\tr((\rho_{x}\rho_{x'}\otimes\openone)M_{c|z})\right]_{x,c,z}^{x'} \\
 & & \left[\tr(\openone\otimes\sigma_y \sigma_{y'})\right]_{y}^{y'} & \left[\tr((\openone\otimes\sigma_{y'})M_{c|z})\right]_{c,z}^{y'} & \left[\tr((\rho_{x}\otimes\sigma_{y'})M_{c|z})\right]_{x,c,z}^{y'} \\
 & & & \left[\tr(M_{c|z}M_{c'|z'})\right]_{c,z}^{c',z'} & \left[\tr((\rho_{x}\otimes\openone)M_{c|z}M_{c'|z'})\right]_{x,c,z}^{c',z'} \\
 & & & & \left[\tr((\rho_{x}\otimes\openone)M_{c|z}M_{c'|z'}(\rho_{x'}\otimes\openone))\right]_{x,c,z}^{x',c',z'}
\end{pmatrix} \ , \nonumber
\end{align} 
where we denote $[M]_i^j$ the blocks $M$ indexed with columns $i$ and rows $j$. Some elements in $\Gamma$ can be readily trivialised. The trace of the $d$-dimensional projectors $\Pi_A$ and $\Pi_B$ can be simplified to $\tr(\Pi_A\otimes\Pi_B)=d^2$. Similarly, normalisation implies $\tr(\rho_{x}\otimes\openone)=\tr(\openone\otimes\sigma_{y})=d$ and $\tr(\rho_{x}\otimes\sigma_{y})=1$, $\forall x,y$. This leaves us with the following moment matrix
\begin{align}
\Gamma = \begin{pmatrix}
d^2 & \left[d\right]_x & \left[d\right]_y & \left[\tr(M_{c|z})\right]_{c,z} & \left[\tr((\rho_{x}\otimes\openone)M_{c|z})\right]_{x,c,z} \\
 & \left[d\tr(\rho_{x}\rho_{x'})\right]_{x}^{x'} & \left[1\right]_{y}^{x'} & \left[\tr((\rho_{x'}\otimes\openone)M_{c|z})\right]_{c,z}^{x'} & \left[\tr((\rho_{x}\rho_{x'}\otimes\openone)M_{c|z})\right]_{x,c,z}^{x'} \\
 & & \left[d\tr(\sigma_y \sigma_{y'})\right]_{y}^{y'} & \left[\tr((\openone\otimes\sigma_{y'})M_{c|z})\right]_{c,z}^{y'} & \left[p(c|xy'z)\right]_{x,c,z}^{y'} \\
 & & & \left[\tr(M_{c|z}M_{c'|z'})\right]_{c,z}^{c',z'} & \left[\tr((\rho_{x}\otimes\openone)M_{c|z}M_{c'|z'})\right]_{x,c,z}^{c',z'} \\
 & & & & \left[\tr((\rho_{x}\otimes\openone)M_{c|z}(\rho_{x'}\otimes\openone)M_{c'|z'})\right]_{x,c,z}^{x',c',z'}
\end{pmatrix} \ . \label{eq:mommat}
\end{align} 
Moreover, we obtain further simplifications of $\Gamma$ by using the orthogonality of the measurement operators $M_{c|z}$; it follows that $\tr(M_{c|z}M_{c'|z})=\delta_{c,c'}\tr(M_{c|z})$. Also, without loss of generality we can take $\rho_x$ and $\sigma_y$ to be pure states, and this implies $\tr(\rho_x\rho_x)=\tr(\rho_x)=1$ and $\tr(\sigma_y\sigma_y)=\tr(\sigma_y)=1$. The remaining elements in $\Gamma$ are left as free variables, constrained only by  the positive semidefiniteness of the moment matrix ($\Gamma\succeq 0$) and the normalisation of the probabilities ($\sum_c p(c|xyz)=1)$, $\forall x,y,z$.

\subsection{Reduction by symmetrization}

Now, we leverage symmetries of the objective function $\mathcal{R}_d$ to further simplify the matrix $\Gamma$. To this end, we observe a number of symmetries in our task, i.e.~specific variable permutations that leave $\mathcal{R}_d$ and the scenario invariant. The advantage of having such symmetries is that we can reduce the number of variables in our problem by considering the average $\Gamma$ compted over the symmetries, namely
\begin{equation}\label{sym}
\Gamma_\text{sym}=\frac{1}{|G|}\sum_{\omega\in G} \omega(\Gamma),
\end{equation}
where $G$ is the set of symmetries and $\sigma$ is their action on the moment matrix. 

We first identify the symmetries in our problem. The winning condition in \eqref{eq:winning_condition} remains unchanged under the following transformations,
\begin{align}
&\text{Sym}_1: (x_0,c)\rightarrow \left\{\begin{matrix}
(x_0+1,c+1) \text{ if } z=d \\
(x_0+1,c-2z) \text{ if } z\neq d 
\end{matrix}\right. &&  \text{Sym}_2: (x_1,c)\rightarrow\left\{\begin{matrix}
(x_1+1,c) \text{ if } z=d \\
(x_1+1,c+1)   \text{ if } z\neq d 
\end{matrix}\right. \\
&\text{Sym}_3: (y_0,c)\rightarrow \left\{\begin{matrix}
(y_0+1,c-1) \text{ if } z=d \\
(y_0+1,c+2z) \text{ if } z\neq d 
\end{matrix}\right. &&  \text{Sym}_4: (y_1,c)\rightarrow \left\{\begin{matrix}
(y_1+1,c) \text{ if } z=d \\
(y_1+1,c+1)   \text{ if } z\neq d 
\end{matrix}\right. \ .
\end{align} 
Note that we associate with $d$ elements  $\text{Sym}_1$  (each element contains the set $c,x,y,z$, and cycles $x_0+1$ in $d$ steps). The analogous holds for  the other three symmetries. To use these to build the symmetrised matrix \eqref{sym}, we will apply the above four symmetries one by one.

To this end,  define a new moment matrix averaged over all elements of $\text{Sym}_1$. That is,
\begin{align}
\bar{\Gamma} = \frac{1}{d} \sum_{\omega\in\text{Sym}_1}  \omega(\Gamma) \ .
\end{align}
To see how the new moment matrix looks like, let us go one by one through all unidentified blocks in $\Gamma$ from \eqref{eq:mommat}.

\textbf{Block $\left[\tr(M_{c|z})\right]_{c,z}$}. The symmetry group $\text{Sym}_1$ rotates the indexes $x_0$ and $c$, covering all of them. Since the elements $\left[\tr(M_{c|z})\right]_{c,z}$ do not depend on $x_0$, we only cycle over $c$ and compute the average. Conveniently, the operators $M_{c|z}$ are normalised, which implies that after averaging over all elements from the symmetry group one ends up with
\begin{align}
	\frac{1}{d} \sum_{\omega\in\text{Sym}_1} \omega(\left[\tr(M_{c|z})\right]_{c,z}) = \frac{1}{d} \left[\tr(\openone\otimes \openone) \right]_{c,z} = \left[d\right]_{c,z} \ .
\end{align}

\textbf{Block $\left[\tr((\openone\otimes\sigma_{y'})M_{c|z})\right]_{c,z}^{y'}$}. The application of the symmetry group $\text{Sym}_1$ will have similar effects on all other elements that do not depend on $x_0$. For instance,
\begin{align}
	\frac{1}{d} \sum_{\omega\in\text{Sym}_1} \omega\left(\left[\tr((\openone\otimes\sigma_{y'})M_{c|z})\right]_{c,z}^{y'}\right) = \frac{1}{d} \left[\tr(\openone\otimes\sigma_{y'})\right]_{c,z}^{y'} = \left[1\right]_{c,z}^{y'} \ .
\end{align}

\textbf{Block $\left[\tr((\rho_{x'}\otimes\openone)M_{c|z})\right]_{c,z}^{x'}$}. Analogously to the previous block, we can apply the third symmetry, $\text{Sym}_3$, leading to the elimination of variables in the following block of $\Gamma$  
\begin{align}
	\frac{1}{d} \sum_{\omega\in\text{Sym}_3} \omega\left(\left[\tr((\rho_{x'}\otimes\openone)M_{c|z})\right]_{c,z}^{x'}\right) = \frac{1}{d} \left[\tr(\rho_{x'}\otimes\openone)\right]_{c,z}^{x'} = \left[1\right]_{c,z}^{x'} \ .
\end{align}

\textbf{Block $\left[\tr(M_{c|z}M_{c'|z'})\right]_{c,z}^{c',z'}$}. We move on applying $\text{Sym}_1$ on the block,
\begin{align}
	\frac{1}{d} \sum_{\omega\in\text{Sym}_1} \omega\left(\left[\tr(M_{c|z}M_{c'|z'})\right]_{c,z}^{c',z'}\right) = \left\{\begin{array}{ll}
	 \frac{1}{d} \left[\tr(\openone\otimes \openone)\right]_{c,z}^{c',z'} = \left[d\right]_{c,z}^{c',z'} & \text{if } z = z', \ c=c' \\
	\left[0\right]_{c,z}^{c',z'} & \text{if } z = z', \ c\neq c' \\
	\left[\alpha_{czc'z'}\right]_{c,z}^{c',z'} & \text{if } z \neq z'
	\end{array}\right. \ .
\end{align}
We find that if $z \neq z'$ the average remains unspecified (thus still an SDP variable which we call $\alpha_{czc'z'}$). However, note that if we apply $\text{Sym}_2$ right after $\text{Sym}_1$, and average over all resulting moment matrices, we can define
\begin{align}
\left[\tilde{\alpha}_{czc'z'}\right]_{c,z}^{c',z'} := \frac{1}{d} \sum_{\omega'\in\text{Sym}_2} \omega'\left(\left[\alpha_{czc'z'}\right]_{c,z}^{c',z'}\right) = \left\{\begin{array}{ll}
\frac{1}{d^2} \sum_{j,l} \left[\tr(M_{c-j2z+l|z}M_{c'-j2z'+l|z'})\right]_{c,z}^{c',z'} & \text{if } z\neq d , \ z'\neq d , \ z \neq z' \\
\frac{1}{d^2} \sum_{j,l} \left[\tr(M_{c+j|z}M_{c'-j2z'+l|z'})\right]_{c,z}^{c',z'} & \text{if } z = d , \ z'\neq d , \ z \neq z'  \\
\frac{1}{d^2} \sum_{j,l} \left[\tr(M_{c-j2z+l|z}M_{c'+j|z'})\right]_{c,z}^{c',z'} & \text{if } z\neq d , \ z' = d , \ z \neq z' 
\end{array}\right. \ .
\end{align}
In all cases, the sub-indexes in all sums become decoupled, and the sums can be re-interpreted to run over all measurement outcomes indexed by $\tilde{c}$ and $\tilde{c}'$ independently. This statement requires an explanation. Consider the first case ($z,z'\neq d$), and let us focus on a fixed element $M_{\tilde{c}|z}$. This element will appear in the sum whenever $\tilde{c} = c-j2z+l$, an equality that can be satisfied $d$ times for all available values of $l$ and $j$ (i.e. for given $l$ one can find only one $j$ that satisfies the equality, and this can be done $d$ times). In all those cases, we find a different element $M_{\tilde{c}'|z'}$ that accompanies the same $M_{\tilde{c}|z}$ for $\tilde{c}'=c-j2z'+l$, given that $z\neq z'$. This means that in the whole sum, we can group up elements as $M_{\tilde{c}|z} \sum_{\tilde{c}'} M_{\tilde{c}'|z'}$. The same applies for the rest of cases, as long as $z\neq z'$. Since we can do that $d$ times for all distinct $M_{\tilde{c}|z}$, we end up with
\begin{align}
\left[\tilde{\alpha}_{czc'z'}\right]_{c,z}^{c',z'} = \frac{1}{d^2}\sum_{\tilde{c}} \sum_{\tilde{c}'} \left[\tr(M_{\tilde{c}|z}M_{\tilde{c}'|z'})\right]_{c,z}^{c',z'} = \frac{1}{d^2}\left[d^2\right]_{c,z}^{c',z'} = \left[1\right]_{c,z}^{c',z'} \ .
\end{align}

\textbf{Block $\left[d\tr(\rho_{x}\rho_{x'})\right]_{x}^{x'}$}. After applying $\text{Sym}_1$ and $\text{Sym}_2$ consecutively, we covered all $x_0$ and $x_1$, which implies that, after averaging, all elements will contain the same members in the sum. Thus,
\begin{align}
	\frac{1}{d^2} \sum_{\omega'\in\text{Sym}_2} \sum_{\omega\in\text{Sym}_1} \omega'\left(\omega\left(\left[d\tr(\rho_{x}\rho_{x'})\right]_{x}^{x'}\right)\right) = \frac{1}{d^2} \sum_{x_0,x_1} \sum_{x'_0,x'_1} \left[d\tr(\rho_{x_0 x_1}\rho_{x'_0 x'_1})\right]_{x}^{x'} =\left\{\begin{array}{ll}
	\left[d\Delta_\rho\right]_{x}^{x'} & \text{if } x\neq x' \\
	\left[d\right]_{x}^{x'} & \text{if } x = x'
	\end{array}\right. \ ,
\end{align}
where we defined $\Delta_\rho:=\frac{1}{d^2}\sum_{xx'}\tr(\rho_x\rho_{x'})$. 

\textbf{Block $\left[ d\tr(\sigma_{y}\sigma_{y'}) \right]_{y}^{y'}$}. Similarly, combining symmetry groups $\text{Sym}_3$ and $\text{Sym}_4$ one finds
\begin{align}
	\frac{1}{d^2} \sum_{\omega'''\in\text{Sym}_4} \sum_{\omega''\in\text{Sym}_3} \omega'''\left(\omega''\left(\left[ d\tr(\sigma_{y}\sigma_{y'}) \right]_{y}^{y'}\right)\right) = \frac{1}{d^2} \sum_{y_0,y_1} \sum_{y'_0,y'_1} \left[d\tr(\sigma_{y_0 y_1}\sigma_{y'_0 y'_1})\right]_{y}^{y'} =\left\{\begin{array}{ll}
	\left[d\Delta_\sigma\right]_{y}^{y'} & \text{if } y\neq y' \\
	\left[d\right]_{y}^{y'} & \text{if } y = y'
	\end{array}\right. \ ,
\end{align}
for $\Delta_\sigma:=\frac{1}{d^2}\sum_{yy'}\tr(\sigma_y\sigma_{y'})$. Both $\Delta_\rho$ and $\Delta_\sigma$ will remain as unspecified SDP variables.

\textbf{Block $\left[\tr((\rho_{x}\rho_{x'}\otimes\openone)M_{c|z})\right]_{x,c,z}^{x'}$}. We continue now applying $\text{Sym}_3$ consecutively after $\text{Sym}_1$ and $\text{Sym}_2$. One then finds that the averages run over all indices $x$, $x'$ and $c$ from the elements 
\begin{align}
	\frac{1}{d^3} \sum_{\omega''\in\text{Sym}_3} \sum_{\omega'\in\text{Sym}_2} \sum_{\omega\in\text{Sym}_1} \omega''\left(\omega'\left(\omega\left(\left[\tr((\rho_{x}\rho_{x'}\otimes\openone)M_{c|z})\right]_{x,c,z}^{x'}\right)\right)\right) =\left\{\begin{array}{ll}
	\left[\Delta_\sigma\right]_{x,c,z}^{x'} & \text{if } x\neq x' \\
	\left[1\right]_{x,c,z}^{x'} & \text{if } x = x'
	\end{array}\right. \ .
\end{align}

\textbf{Block $\left[\tr((\rho_{x}\otimes\openone)M_{c|z}M_{c'|z'})\right]_{x,c,z}^{c',z'}$}. Still only applying $\text{Sym}_3$ consecutively after $\text{Sym}_1$ and $\text{Sym}_2$, once the average runs over all $x$, $c$ and $c'$ indices, we find
\begin{align}
	\frac{1}{d^3} \sum_{\omega''\in\text{Sym}_3} \sum_{\omega'\in\text{Sym}_2} \sum_{\omega\in\text{Sym}_1} \omega''\left(\omega'\left(\omega\left(\left[\tr((\rho_{x}\otimes\openone)M_{c|z}M_{c'|z'})\right]_{x,c,z}^{c',z'}\right)\right)\right) =\left\{\begin{array}{ll}
	\left[\frac{1}{d}\right]_{x,c,z}^{x'} & \text{if } z\neq z' \\
	\left[1\right]_{x,c,z}^{x'} & \text{if } z = z' , \ c = c' \\
	\left[0\right]_{x,c,z}^{x'} & \text{if } z = z' , \ c \neq c'
	\end{array}\right. \ .
\end{align}

\textbf{Block $\left[\tr((\rho_{x}\otimes\openone)M_{c|z}M_{c'|z'}(\rho_{x'}\otimes\openone))\right]_{x,c,z}^{x',c',z'}$}. Finally, we apply $\text{Sym}_4$ and average over all resulting elements. Once the average runs over all $x$, $c$ and $c'$ indices, we find
\begin{align}
	\frac{1}{d^4} \sum_{\omega\in\text{All Syms}} \omega\left(\left[\tr((\rho_{x}\otimes\openone)M_{c|z}M_{c'|z'}(\rho_{x'}\otimes\openone))\right]_{x,c,z}^{x',c',z'}\right) =\left\{\begin{array}{l}
	\left[0\right]_{x,c,z}^{x',c',z'} \ \text{if } z = z' , \ c \neq c' \\
	\left[1\right]_{x,c,z}^{x',c',z'} \ \text{if } z = z' , \ c = c', \ x = x' \\
	\left[\Delta_\rho\right]_{x,c,z}^{x',c',z'} \ \text{if } z = z' , \ c = c', \ x \neq x' \\
	\left[\frac{1}{d}\right]_{x,c,z}^{x',c',z'} \ \text{if } z \neq z', \ x = x' \\
	\left[\frac{\Delta_\rho}{d}\right]_{x,c,z}^{x',c',z'} \ \text{if } z \neq z', \ x \neq x' 
	\end{array}\right. \ .
\end{align}
The first case leading to $\left[0\right]_{x,c,z}^{x',c',z'}$ is trivialized by considering $M_{c|z}$ orthogonal projectors. The second and third cases reduce this block to $\left[\tr((\rho_{x}\rho_{x'}\otimes\openone)M_{c|z})\right]_{x,c,z}^{x'}$ previously considered. The forth case reduces this block to $\left[\tr((\rho_{x}\otimes\openone)M_{c|z}M_{c'|z'})\right]_{x,c,z}^{c',z'}$ also considered above. Lastly, in the $z \neq z'$, $x \neq x'$ case, averaging over all elements in the symmetries $\text{Sym}_3$ and $\text{Sym}_4$ allows $c$ and $c'$ to cycle independently, resulting in $\frac{1}{d^2}\left[\tr(\rho_{x}\rho_{x'}\otimes\openone)\right]_{x,c,z}^{x',c',z'}$. Hence, after performing the rest of the symmetries, one recovers $\left[\frac{\Delta_\rho}{d}\right]_{x,c,z}^{x',c',z'}$.

\textbf{Block $\left[\tr(\rho_{x}\otimes\sigma_y)M_{c|z}\right]_{x,c,z}^{y}$}. Note that the elements from the moment matrix containing the observable probabilities do not result in identifiable values after averaging over all symmetries. This is because all indices $x_0,x_1,y_0,y_1,c,z$ appear in those elements, which makes that e.g.~the indices $c$ cannot be summed over independently. However, averaging over the identified symmetries, one finds 
\begin{align}\label{eq:avg_probs}
	\frac{1}{d^4} \sum_{\omega\in\text{All Syms}} \omega\left(\left[\tr(\rho_{x}\otimes\sigma_{y})M_{c|z}\right]_{x,c,z}^{y}\right) =\left\{\begin{array}{ll} 
	 \left[\displaystyle\frac{1}{d^4} \displaystyle\sum_{ijkl} p(c+j+l-2z(i-k)|x_0+i, x_1+j, y_0+k, y_1+l, z)\right]_{x,c,z}^{y} & \text{if} \ z\neq d \\
\left[\displaystyle\frac{1}{d^4} \displaystyle\sum_{ijkl} p(c+i-k|x_0+i, x_1+j, y_0+k, y_1+l, z)\right]_{x,c,z}^{y} & \text{if} \ z=d 
	\end{array}\right. \ .
\end{align}
Note that all probabilities containing the winning event $c=w_z(x,y)$ will be inside the same average since $c'=w_{z'}(x',y')$ for $x_0'=x_0+i$, $x_1'=x_1+j$, $y_0'=y_0+k$, $y_1'=y_1+l$ and $c'=c+j+l-2z(i-k)$. Also, the average containing all probabilities corresponding to the event of missing the winning condition by $m$ (i.e.~$c=w_z(x,y)+m$) will be grouped up in the same averages as well. Doing so, we can identify $d(d+1)$ distinct averages (for each output $c$ and input $z$). To this end, we define
\begin{align}
q(\tilde{c}_z|z) :=& \frac{1}{d^4} \sum_{x_0, x_1, y_0, y_1}  p(c-w_z(x,y)|x_0 x_1 y_0 y_1 z) \ .
\end{align}
We group all equivalent averages from \eqref{eq:avg_probs} in each $q(\tilde{c}_z|z)$ for all distinct $c$ and $z$. These elements are conveniently defied to collect all distinct averages resulting from the block $\left[\tr(\rho_{x}\otimes\sigma_y)M_{c|z}\right]_{x,c,z}^{y}$ containing events that miss the winning condition by $\tilde{c}_z:=c-w_z(x,y)$. Physically, this re-naming can be interpreted as a classical post-processing of labelling events. Consequently, note that all winning conditions are grouped up whenever $\tilde{c}_z=0$. Hence, the averaged winning condition becomes
\begin{align}
\mathcal{R}_d = \frac{1}{d+1}\sum_{z=0}^{d}q(0|z) \ .
\end{align}

At the end of the day we end up with a total of $d(d+1) + 2$ (that is all $q(\tilde{c}_z|z)$ plus $\Delta_\rho$ and $\Delta_\sigma$) unidentified elements in the moment matrix taken as SDP variables. 
This drastic reduction allows us to re-consider the size of the original moment matrix, as now we know most of its elements. In fact, consider the minor $G$ of $\Gamma$ containing only the following elements
\begin{align}
G = \begin{pmatrix}
\left[d\tr(\sigma_y \sigma_{y'})\right]_{y}^{y'} & \left[p(c|xy'z)\right]_{x,c,z}^{y'} \\
\left[p(c|xy'z)\right]^{x',c',z'}_{y} & \left[\tr((\rho_{x}\otimes\openone)M_{c|z}(\rho_{x'}\otimes\openone)M_{c'|z'})\right]_{x,c,z}^{x',c',z'}
\end{pmatrix} \ . \nonumber
\end{align} 
We know that, after the whole symmetrization process, we can re-consdier the averaged $G$ over all symmetry groups, i.e.
\begin{align}
\bar{G} = \begin{pmatrix}
\bar{G}_{\sigma\sigma} & \bar{G}_{\sigma M\rho} \\
\bar{G}_{\sigma M\rho}^T & \bar{G}_{M\rho M\rho}
\end{pmatrix} \ , 
\end{align} 
for 
\begin{align}
&\bar{G}_{\sigma \sigma} =\left\{\begin{array}{ll} 
	\left[d\Delta_\sigma\right]_{y}^{y'} & \text{if } y\neq y' \\
	\left[d\right]_{y}^{y'} & \text{if } y = y'
	\end{array}\right. \ , \quad \bar{G}_{\sigma M\rho} =
	\left[q(c-w_z(x,y')|z)\right]_{x,c,z}^{y'} \ , \quad \bar{G}_{M\rho M\rho} =\left\{\begin{array}{l}
	\left[0\right]_{x,c,z}^{x',c',z'} \ \text{if } z = z' , \ c \neq c' \\
	\left[1\right]_{x,c,z}^{x',c',z'} \ \text{if } z = z' , \ c = c', \ x = x' \\
	\left[\Delta_\rho\right]_{x,c,z}^{x',c',z'} \ \text{if } z = z' , \ c = c', \ x \neq x' \\
	\left[\frac{1}{d}\right]_{x,c,z}^{x',c',z'} \ \text{if } z \neq z', \ x = x' \\
	\left[\frac{\Delta_\rho}{d}\right]_{x,c,z}^{x',c',z'} \ \text{if } z \neq z', \ x \neq x' 
	\end{array}\right. \ .
\end{align}
Replacing the original moment matrix $\bar{\Gamma}$ with $\bar{G}$ still provides with a valid solution to our problem, since $\bar{\Gamma} \succeq 0 \Rightarrow \bar{G} \succeq 0$. Hence, we will only consider the symmetrized minor $\bar{G}$.

\subsection{Block-diagonalization}

Although this SDP form is already greatly simplified, the size of $\bar{G}$ still scales rapidly with growing $d$, which makes the constraint $\bar{G} \succeq 0$ expensive to compute. To completely exploit the symmetries in our problem, we reduce the SDP complexity by performing a block-diagonalisation. Here, we provide an overview of the core idea and outline how we utilize existing techniques for simultaneous block-diagonalization in our approach. For a more detailed explanation, we direct interested readers to Ref.~\cite{murota2010}.

Consider the vectorized form of our SDP problem, written as
\begin{align}
	\underset{\Gamma}{\text{maximize}} & \quad \tr\left[\Gamma A\right] \\
	\text{subject to} & \quad \tr\left[\Gamma B_{xyz}\right] = 1 \ , \quad \Gamma \succeq 0 \ . \nonumber
\end{align}
Here, $A$ is a matrix with $\frac{1}{d^4 (d+1)}$ in the positions where $\Gamma$ contains the elements $p(c=w_z|xyz)$, and $B_{xyz}$ another matrix with ones in the positions where $\Gamma$ contains $p(c|xyz)$ for all $c$, and zeros otherwise. After performing all symmetrization procedures and reductions mentioned above, we can re-write our SDP as
\begin{align}
	\underset{\bar{G}}{\text{maximize}} & \quad \tr\left[\bar{G} \bar{A}\right] \\
	\text{subject to} & \quad \tr\left[\bar{G} \bar{B}_{z}\right] = 1 \ , \quad \bar{G} \succeq 0 \ . \nonumber
\end{align}
Here, $\bar{A}$ is a matrix with $\frac{1}{(d+1)}$ in the positions where $\bar{G}$ contains the elements $q(0|z)$ for all $z$, and zeros otherwise; and $\bar{B}_{z}$ with ones in the positions where $\bar{G}$ contains $q(\tilde{c}_z|z)$ for all $\tilde{c}_z$, and zeros otherwise. The moment matrix $\bar{G}$ contains a significant amount of elements whose values are known or have a fixed structure, leaving only $d(d+1)+2$ unknown values as SDP variables. Let $\{x_i\}$ denote the set of distinct elements in our moment matrix $\bar{G}$. In our specific case we have
\begin{align}
	\{x_i\}=\left\{d,1,0,\Delta_\rho,d\Delta_\sigma,\frac{\Delta_\rho}{d},\frac{1}{d},\{q(\tilde{c}_z|z)\}_{\tilde{c}_z,z}\right\} \ .
\end{align}
Let $X_{x_i}$ be a matrix with ones in the slots where $\bar{G}$ contains the elements  $x_i$, and zeros otherwise. This way, $\bar{G}=\sum_i x_i X_{x_i}$. Note that now $\bar{A}$ is essentially $\frac{1}{d+1} \sum_z X_{q(0|z)}$. To find a block-diagonal form of our SDP, we need to find a unitary transformation $T$ that brings the complete set of $\{X_{x_i}\}$ simultaneously into a block-diagonal form. That is,
\begin{align}
T^{\dagger}\bar{G}T = \sum_i x_i T^\dagger X_{x_i} T \quad \text{for} \quad T^\dagger X_{x_i} T = \begin{pmatrix}
\tilde{X}_{x_i}^{(1)} & 0 & \cdots \\
0 & \tilde{X}_{x_i}^{(2)} & \cdots \\
\vdots & \vdots & \ddots
\end{pmatrix} = \underset{l}{\bigoplus} \ \tilde{X}_{x_i}^{(l)} \ .
\end{align}
Then we can re-write our SDP in the block-diagonal form
\begin{align}
	\underset{\Delta_\rho,\Delta_\sigma,\{q(\tilde{c}_z|z)\}}{\text{maximize}} & \quad \frac{1}{d+1}\sum_{z}q(0|z) \\
	\text{subject to} & \quad \sum_i x_i \tilde{X}_{x_i}^{(l)} \succeq 0 \ \forall l , \quad \sum_{\tilde{c}_{z}}q(\tilde{c}_z|z) = 1 \ \forall z \ . \nonumber
\end{align}
Our SDP has been reduced from restricting the positive-semidefinite condition $\bar{G} \succeq 0$ of a significantly large matrix, to the set of constraints $\sum_i x_i \tilde{X}_{x_i}^{(l)} \succeq 0$ of matrices with much reduced sizes. In order to find the transformation $T$ for block-diagonalisation, we employ an algorithm developed in Ref.~\cite{zhang2021} which consists in finding the eigendecomposition of a randomized convex combination of the set of matrices $\{X_{x_i}\}$, inspired in Ref.~\cite{murota2010}.

With the presented symmetrization and block-diagonalization techniques, we are able to compute the SDP to prove the principal conjecture in the main text in dimensions $d=3,5,7$.

\subsection{Dual SDP}
Finally we derive the dual SDP. This allows us to obtain an analytical certificate of the bounds computed through the primal SDP. Let us consider again the primal formulation of the reduced SDP,
\begin{align}
\underset{\Delta_\rho,\Delta_\sigma,\{q(\tilde{c}_z|z)\}}{\text{maximize}} & \quad \frac{1}{d+1}\sum_{z}q(0|z) \\
\text{subject to} & \quad d\tilde{X}_d^{(l)} + \tilde{X}_1^{(l)} + \Delta_\rho \tilde{X}_{\Delta_\rho}^{(l)} + d\Delta_\sigma \tilde{X}_{d\Delta_\sigma}^{(l)} + \frac{\Delta_\rho}{d}\tilde{X}_{\frac{\Delta_\rho}{d}}^{(l)} + \frac{1}{d}\tilde{X}_{\frac{1}{d}}^{(l)} + \sum_{\tilde{c}_z,z}q(\tilde{c}_{z}|z)\tilde{X}^{(l)}_{q(\tilde{c}_{z}|z)} \succeq 0 \ \forall l \nonumber \\
& \quad \sum_{\tilde{c}_{z}}q(\tilde{c}_z|z) = 1, \forall z \ . \nonumber
\end{align}
For both constraints of the primal SDP, we incorporate the dual positive-semidefinite variables $Y^{(l)}$ corresponding to the first constraint, and $\nu_z$ for the normalisation constraint. The Lagrangian reads,
\begin{align}
\mathcal{L} &= \frac{1}{d+1}\sum_{z}q(0|z) + \sum_l \tr\left[ Y^{(l)} \left( d\tilde{X}_d^{(l)} + \tilde{X}_1^{(l)} + \Delta_\rho \tilde{X}_{\Delta_\rho}^{(l)} + d\Delta_\sigma \tilde{X}_{d\Delta_\sigma}^{(l)} + \frac{\Delta_\rho}{d}\tilde{X}_{\frac{\Delta_\rho}{d}}^{(l)} + \frac{1}{d}\tilde{X}_{\frac{1}{d}}^{(l)} + \sum_{\tilde{c}_z,z}q(\tilde{c}_{z}|z)\tilde{X}^{(l)}_{q(\tilde{c}_{z}|z)} \right) \right] \\
&+ \sum_z \nu_z \left(1-\sum_{\tilde{c}_{z}}q(\tilde{c}_z|z)\right) \nonumber \\
&= \sum_{\tilde{c}_z,z}q(\tilde{c}_z|z)\left\{ \frac{\delta_{\tilde{c}_z,0}}{d+1} + \sum_l \tr\left[Y^{(l)}\tilde{X}^{(l)}_{q(\tilde{c}_{z}|z)}\right] - \nu_z \right\} + \Delta_\rho \sum_l \tr\left[ Y^{(l)} \left( \tilde{X}_{\Delta_\rho}^{(l)} + \frac{1}{d}\tilde{X}_{\frac{\Delta_\rho}{d}}^{(l)} \right) \right] + \Delta_\sigma \sum_l \tr\left[ Y^{(l)} d \tilde{X}_{d\Delta_\sigma}^{(l)} \right] \nonumber \\
& + \sum_l \tr\left[ Y^{(l)} \left( d\tilde{X}_d^{(l)} + \tilde{X}_1^{(l)} + \frac{1}{d}\tilde{X}_{\frac{1}{d}}^{(l)} \right) \right] + \sum_z \nu_z  \nonumber
\end{align}
Given a solution of the primal SDP, the three first terms of the Lagrangian will vanish. Therefore, the remaining terms will yield an upper bound on the primal objective function. The dual can therefore be formulated as a minimization of the supremum of the Lagrangian over the dual variables. However, this minimisation will diverge unless all terms accompanying each primal variable are nullified. These will define the constraints that the dual variables are required to fulfil. The dual SDP can therefore be written as
\begin{align}
\underset{Y^{(l)},\nu_z}{\text{minimize}} & \quad \sum_l \tr\left[ Y^{(l)} \left( d\tilde{X}_d^{(l)} + \tilde{X}_1^{(l)} + \frac{1}{d}\tilde{X}_{\frac{1}{d}}^{(l)} \right) \right] + \sum_z \nu_z \\
\text{subject to} & \quad  \frac{\delta_{\tilde{c}_z,0}}{d+1} + \sum_l \tr\left[Y^{(l)}\tilde{X}^{(l)}_{q(\tilde{c}_{z}|z)}\right] - \nu_z = 0 \ \forall \tilde{c}_z,z \nonumber \\
& \quad \sum_l \tr\left[ Y^{(l)} \left( \tilde{X}_{\Delta_\rho}^{(l)} + \frac{1}{d}\tilde{X}_{\frac{\Delta_\rho}{d}}^{(l)} \right) \right] = 0 \nonumber \\
& \quad \sum_l \tr\left[ Y^{(l)} d \tilde{X}_{d\Delta_\sigma}^{(l)} \right] = 0 \nonumber \\
& \quad Y^{(l)} \succeq 0 \ \forall l \nonumber \ .
\end{align}
Any feasible point on the dual SDP represents an upper bound on the averaged winning condition $\mathcal{R}_d$ from the task presented in the main text. We computed both the dual and primal SDPs for dimensions $d=3,5,7$. In all cases, we found matching results, proving the conjecture in the main text.


\end{widetext}

\end{document}